# All-optical reversible manipulation of exciton and trion emissions in monolayer $WS_2$

Chaoli Yang,[1,2] Yan Gao,[1,2] Chengbing Qin,[1,2*] Xilong Liang,[1,2] Shuangping Han,[1,2] Guofeng Zhang,[1,2] Ruiyun Chen,[1,2] Jianyong Hu,[1,2] Liantuan Xiao,[1,2*] and Suotang Jia[1,2]

[1] State Key Laboratory of Quantum Optics and Quantum Optics Devices, Institute of Laser Spectroscopy, Shanxi University, Taiyuan, Shanxi 030006, China.

[2] Collaborative Innovation Center of Extreme Optics, Shanxi University, Taiyuan, Shanxi 030006, China.

[*]Author to whom correspondence should be addressed.

E-mail:

Chengbing Qin, chbqin@sxu.edu.cn

Liantuan Xiao, xlt@sxu.edu.cn

**Abstract**

Monolayer transition metal dichalcogenides (TMDs) are direct gap semiconductors emerging promising applications in diverse optoelectronic devices. To improve performance, recent investigations have been systematically focused on the tuning of their optical properties. However, an all-optical approach with the reversible feature is still a challenge. Here we demonstrate the tunability of the photoluminescence (PL) properties of monolayer $WS_2$ via laser irradiation. The modulation of PL intensity, as well as the conversion between neutral exciton and charged trion have been readily and reversibly achieved by using different laser power densities. We attribute the reversible manipulation to the laser-assisted adsorption and desorption of gas molecules, which will deplete or release free electrons from the surface of $WS_2$ and thus modify its PL properties. This all-optical manipulation, with advantages of reversibility, quantitative control, and high spatial resolution, suggests promising applications of TMDs monolayers in optoelectronic and nanophotonic applications, such as optical data storage, micropatterning, and display.

## 1. Introduction

Atomically thin two-dimensional (2D) transition metal dichalcogenides (TMDs) with the chemical formula $MX_2$ (M=Mo, W, and X=S, Se) have attracted great interest recently, due to their unique electric and optical properties as well as their potential applications in diverse optoelectronic devices.[1,2] Compared to their bulk crystals and the multilayer form, monolayer $MX_2$ are direct gap semiconductors and manifest bright photoluminescence (PL) in the visible region even at room temperature.[3] Due to the spatial confinement of electron motion and reduced Coulomb screening in the 2D structures, monolayer $MX_2$ are also emerging as a new platform for exploring many-body interactions, such as Auger recombination and exciton-exciton annihilation.[4-7] However, on the other hand, the large surface-to-volume ratio of monolayer $MX_2$ makes them extremely sensitive to the changes of their surrounding environments, such as gating,[8] doping,[9] defects,[10] temperature,[11] excitation power[12] and collection locations.[13] Thus, exploring and manipulating the optical properties of these monolayer $MX_2$ under different conditions are vital for the fundamental physics and optoelectronic applications.

Recently, unprecedented tunability of the optical properties of layered $MX_2$ have been achieved both theoretically and experimentally. For example, by spinning coating p/n-type organic dopants (for instance, TCNQ and NADH molecules) or even biomolecules (such as DNA nucleobases) on monolayer $MX_2$,[9,14-16] the conversion among neutral exciton and positive or negative trion can be undoubtedly determined, although these manipulations are irreversible. The reversible electrostatic tunability of the exciton charging effects in monolayer $MX_2$ has been realized by using field-effect transistor (FET) structure or molecular physisorption gating.[8,17,18] However, the contaminants and impurities are always inevitably introduced during the fabrication of the FET structure, which may significantly reduce the performance of optoelectronic devices. It can be partly avoided by using molecular physisorption gating, while the vacuum conditions and pumping apparatus impede their extensive applications.[18] Hence, an all-optical scheme that can readily operate in the atmosphere and reversibly manipulates the optical properties of monolayer $MX_2$ without introducing extra impurities is highly desired.

Here, we show that the switch between exciton and trion of monolayer $WS_2$, and thus the modulation of its PL can be reversibly fulfilled by changing the power density of the irradiation laser. Under the high power density, excitons will convert into trions, resulting in the quenching of PL intensity and redshift of its spectra. After quenching, the exciton can be recovered from trions by laser irradiation with a lower power density. Consequently, the manipulation of excitons and trions, PL intensity, as well as PL spectra, can be reversibly manipulated.

## 2. Experimental section

The optical experiments, including laser irradiation and PL collection, are performed by using a home-built scanning confocal microscope. The experimental schematic has been described in detail elsewhere,[19,20] which can also be found in the Electronic Supplementary Information (ESI, Fig. S1). Particularly, a 532 nm continuum-wave (CW) laser was used as both the irradiation source to modify the $WS_2$ and the excitation laser to characterize its optical properties (PL imaging). The laser was focused by a dry objective (Nikon, 100×, NA=0.9) with the lateral dimension of the focal spot about 1 μm. The $WS_2$ sample was placed on a motorized three-dimensional piezoelectric translation nano-stage (PZT, Tritor, 200/20SG) with typical repeatability less than 35 nm. The high spatial resolution guarantees that the laser-focused position on the $WS_2$ sample during cycle irradiation is the same. PL imaging of monolayer $WS_2$ was created by moving the sample with respect to the focused laser beam in a programmable and control way. PL intensity was collected by using the same objective. After passing through a dichroic mirror (Semrock, Di01-R532-25x36) and a long-pass filter (Semrock, NF01-532U-25) to block the back scatted laser as well as the background noise, PL was further filtered spatially using a 100 μm pinhole and then divided into two beams through a beam splitter. The PL intensity and its spectrum were synchronously recorded, by using a single-photon detector (PerkinElmer, SPCM-AQR-15) and a monochromator equipped with a cooled charge-coupled device (CCD, Princeton Instrument, PIXIS), respectively. The incident laser power used in this experiment was ranging from 0.1 mW to 10 mW, calibrated using a power meter (Nova II, P/N7Z01550) before each measurement.

## 3. Results and discussions

3.1 *Sample characterizations*

In the experiment, n-type $WS_2$ was first grown on a sapphire substrate by using the conventional chemical vapor deposition (CVD) method. Considering that the switch between excitons and trions of monolayer $MX_2$ on $SiO_2$/Si substrate was more pronounced,[21], we then transferred the $WS_2$ to a $SiO_2$/Si substrate with a poly(methyl methacrylate) (PMMA) assisted method.[22] Fig. 1(a) presents an optical image of a typical sample. Isolated flakes with a triangular shape and edge lengths ranging from 10 to 20 μm can be clearly observed. The thickness of these $WS_2$ flakes has been explored by atomic force microscopy (AFM) characterization, as shown in Fig. 1(b). The thickness of ~0.8 nm (the cross section height is presented in the inset) confirms that the flake is a monolayer.[23] The relatively homogeneous color contrast in the AFM image also indicates that the basal plane of the monolayer $WS_2$ is flat and uniform. We additionally note that the measured thickness is in agreement with the result of the Raman spectroscopy, as displayed in Fig. 1(c).

Raman spectroscopy was performed by a Horiba LabRAM HR Raman microscope using the 532 nm laser excitation. The 521 cm$^{-1}$ phonon mode from the Si substrate was used for calibration. As presented in Fig. 1(c), the peaks located at 417 cm$^{-1}$ and 356 cm$^{-1}$ can be attributed to the first order of the out-of-plane $A_{1g}(\Gamma)$ and in-plane $E^1_{2g}(\Gamma)$ modes of $WS_2$. A frequency difference of 61 cm$^{-1}$ between the $A_{1g}(\Gamma)$ and $E^1_{2g}(\Gamma)$ modes is definitely determined. These signatures agree well with previous Raman studies of monolayer $WS_2$.[24,25] Furthermore, the multi-peak Lorentzian fits shown in Fig. 1(c) clearly reveals that some high order vibrational modes are contributed to the Raman signal as well. As it is widely accepted, the broadening and strong intensity of $E^1_{2g}(\Gamma)$ peak can be attributed to a second-order longitudinal acoustic phonon,[26] 2LA(M) mode, emerging at 351 cm$^{-1}$. An even weak and lower peak (328 cm$^{-1}$) has been attributed to the 2LA(M)-2E$^1_{2g}$(M) process, a high frequency mode beyond Si located at 583 cm$^{-1}$ is generally attributed to A$_{1g}$(M)+LA(M).[23,27] Here, E$^1_{2g}$(M) and A$_{1g}$(M) are the in-plane and out-of-plane vibrational modes related to the phonon dispersion near the M point of the Brillouin zone.

As expected, strong and uniform PL is observed at room temperature over the monolayer triangular shape, with the peak near 1.96 eV, as shown in Fig. 1(d). The peak can be well fitted by two Lorentzian functions. Generally, the A excitons are strong in monolayer MX$_2$ due to the reduced dielectric screening. Furthermore, the difference between the fitted two peaks is about 40 meV, coinciding with the binding energy of trion reported.[4] Thus, the higher energy peak at 1.98 eV can be attributed to the neutral exciton emission (A$^0$), and the lower peak at 1.94 eV to the negative charged trion emission (A$^-$).[4,28] Here, the trion is formed by an electron binding with an exciton, due to the n-type feature of monolayer WS$_2$ and the extra free electrons on its surface. The formation of trions will suppress the radiative recombination of neutral excitons and thus reduce PL intensity as well as change the emission energy. Therefore, the PL can be manipulated by switching the ratio of excitons and trions. Hereinafter, we show that this manipulation can be readily and reversibly achieved by using laser irradiation with different power densities.

3.2 *All-optical reversible manipulation of the exciton and trion emissions*

Fig. 2(a) depicts PL intensity of monolayer WS$_2$ varied as laser irradiation with different power densities. At the initial stage with the low power excitation ($P_1$=20 kW/cm$^2$), the absence of changes in both PL intensity ($t_0$ to $t_1$ in the figure) and spectra (see Fig. S2 in ESI) hints that the crystal structure and optical properties of monolayer WS$_2$ are maintained during this stage. However, an unanticipated PL quenching emerged when the power density was switched to high ($P_2$=900 kW/cm$^2$). This quenching is a time-consuming process (typically more than 500 s to stable, see Fig. S3 in ESI), rather than instantaneous. To understand this phenomenon, we switched off the laser for a duration (about 50s from $t_2$ to $t_3$) during the quenching process and then switched on the laser again, as shown in Fig. 2(a). Comparing the PL intensity before laser off ($t_2$) and after laser on ($t_3$), we note that no significant changes occur. Thus, we suggest that the thermal effect because of the strong excitation laser can be excluded in this process. Due to the quenching under high power excitation, PL will become weaker compared with the initial stage, as shown in Fig. 2(b) and 2(c) (PL imaging were performed under the same laser excitation). Amazingly, PL can be restored to the initial intensity by laser irradiation with an appropriate power density (as $P_1$ used here and the details will be discussed below), as presented in Fig. 2(d) and 2(e), respectively. This restoration is robust and stable, as the $t_6$ and $t_7$ shown in Fig. 2. Undoubtedly, an all-optical reversible manipulation on the PL intensity of monolayer WS$_2$ has been fulfilled. Compared with previous works that modified PL of layered MX$_2$ by thermal annealing,[11,29] molecular doping[9,16], and physisorption gating,[18] our all-optical manipulation not only holds the merits of reversibility but also maintains the high spatial resolution (limited by the diffraction). In particular, this high spatial resolution provides promising applications in optical data storage, display technology, and relevant optoelectronic devices.[30,31]

To elucidate the underlying physical mechanisms relevant to the reversible manipulation, we conducted PL spectra during the quenching process under high power excitation and the recovery process under low power excitation, respectively. Fig. 3(a) clearly demonstrates that the intensity of PL spectra decreases gradually as a function of laser irradiation duration. Meanwhile, the full width at half maximum (FWHM), and the peak position of the PL spectra also emerge slight changes during the quenching process. To interpret these spectral changes, all the spectral profiles are decomposed into two Lorentz peaks. As presented in Fig. 3(b), with the increasing duration of laser irradiation, the relative proportion of $A^0$ decreases gradually, while that of $A^-$ increases. By dividing the integral intensity of each component to the full spectra, the spectral weight of both $A^0$ and $A^-$ can be determined, as shown in Fig. 3(c). Note that, the weight of $A^0$ converts into $A^-$ rapidly at the beginning of the quenching process, and then becomes slowly and trends to stable. The conversion rate can be determined empirically by fitting the two traces with a single exponential function,

$$W_{A^0/A^-}(t) = W_{A^0/A^-}(0) + a_{A^0/A^-} \times \exp(k_{A^0/A^-} \cdot t) \tag{1}$$

Here, $W_{A0/A-}(t)$ is the spectral weight of $A^0$ or $A^-$ at time $t$, $k_{A0}$ ($k_{A-}$) is the rate of $A^0$ ($A^-$) converting into $A^-$ ($A^0$). Based on the fitted curves (as the solid lines shown in Fig. 3(c)), $k_{A0}$ and $k_{A-}$ are determined to be $5.8 \times 10^{-2}$ $s^{-1}$, respectively. Furthermore, we note that the peak position of $A^0$ redshifts from 1.963 eV to 1.960 eV in this process, and that of $A^-$ redshifts from 1.927 eV to 1.924 eV, respectively, as shown in Fig. 3(d). Nevertheless, their energy differences, $\Delta(E_{A0}-E_{A^-})$, namely the binding energy of trion, remain almost unchanged during this process and agree well with reported values of about 34 meV.[32,33] Thus, we can conclude that the quenching of PL is originating from the conversion from $A^0$ to $A^-$.

We also studied the time evolution of PL spectra in the recovery process after quenching, which confirms that the restoration of PL results from the conversion from $A^-$ to $A^0$, as presented in Fig. 4. In this stage, PL spectra increase gradually, accompanied by characteristic changes of the emission profiles. Based on the Lorentzian fits shown in Fig. 4(b), the increase of $A^0$ component can be clearly determined. One may find that a significant difference is present between the spectral profile for the end of the quenching process (*i.e.,* the last spectrum shown in Fig. 3(b)) and the beginning of the recovery process (*i.e.,* the first spectrum shown in Fig. 4(b)). This can be understood as the $A^-$ component dominates the spectra at the excitation with low power densities. The spectral weights of $A^0$ and $A^-$ as a function of laser irradiation duration in this stage have been presented in Fig. 4(c), from which the conversion rates of $k'_{A0}$ and $k'_{A^-}$ are determined to be $7.7 \times 10^{-3}$ s$^{-1}$ by equation 1. That's to say, the conversion rate from $A^-$ to $A^0$ in the recovery process is almost one order of magnitude slower than the conversion rate from $A^0$ to $A^-$ in the quenching process ($5.8 \times 10^{-2}$ s$^{-1}$). As expected, the peak positions of both $A^0$ and $A^-$ are blue-shift in this stage. Additionally, both the peak position and binding energy of trions are higher than that in the quenching process with the high power excitation. This result has been speculated to the band structure renormalization under the high excitation power.[12]

To gain further insight into the nature of this effect, we performed this reversible manipulation under various laser power densities. As shown in Fig. 5(a), the quenching process becomes weaker as the irradiation power density decreases from 900 kW/cm$^2$ to 750 kW/cm$^2$. This process will be even absent when the power is lower than ~400 kW/cm$^2$. Conversely, the behaviors in the recovery process are more plentiful. In the relatively low irradiation power density (such as 10 kW/cm$^2$ shown in Fig. 5(b)), PL will be enhanced slightly but pretty slowly, rather than restored to the initial intensity. As the power density of irradiation laser increases, this situation will improve gradually and even reach the initial intensity under appropriate power density, such as 20 kW/cm$^2$ shown in Fig. 5(b). However, PL may be maintained or even quenched mildly when the irradiation power density further increases, as 25 kW/cm$^2$ and 30 kW/cm$^2$ used in the experiments.

3.3 *Proposed mechanism to explore the manipulation*

Here we propose a potential mechanism to explore the observed results. Laser irradiation with strong power density may result in the structural degradation of layered 2D materials and synchronously modify their optical properties, such as PL enhancement and energy shifts.[12,34,35] However, this kind of manipulation is destructive and non-reversible, and thus it can be excluded here. Previous works have also reported that PL quantum yield of monolayer MoS$_2$ can be availably modulated by physical adsorption/desorption of O$_2$ and/or H$_2$O molecules through pumping up or down the gas pressure.[18] This reversible manipulation can provide orders of magnitude broader control of PL emission. Inspired by this result, we attribute our reversible manipulation to the laser-assisted adsorption/desorption of gas molecules (such as $O_2$ and $H_2O$) on the surface of monolayer $WS_2$, as illustrated schematically in Fig. 5(c). Here we assume that (i) activation energy is required for both adsorption ($\Delta E_{ad}$) and desorption processes ($\Delta E_{de}$). Furthermore, the activation energy of the desorption process is higher than that of the adsorption process ($\Delta E_{de} > \Delta E_{ad}$). (ii) We also propose that laser irradiation will heat the layered samples and thus provide the activation energy to adsorb or desorb the gas molecules. In our experiment, n-type $WS_2$ is used, and thus abundant free electrons are available on its surface. Although activation energy is required, gas molecules will still accumulate on the surface of monolayer $WS_2$ and reach saturation after exposing in the air for a long period (the sample has been exposed in air for months before performing this experiment), due to that the activation energy of desorption is higher than that of the adsorption process. According to the previous works,[18] we can conclude that the adsorption of gas molecules will deplete the free electrons, while desorption will increase. Consequently, the amounts of neutral excitons and negative trions can be switched by these two processes.

(I) At the initial stage, monolayer $WS_2$ is irradiated by the relatively low power density. Hence the laser-induced heat effect is not pronounced. Therefore, the activation energy is not high enough for desorbing gas molecules from the surface. Although adsorption may occur during this stage, the molecule concentration has already reached saturation before laser irradiation. Consequently, no more gas molecules can be adsorbed, and PL of monolayer $WS_2$ will maintain, as $t_0$ to $t_1$ shown in Fig.

2(a).

(II) When the activation energy for desorbing gas molecules has reached by laser irradiation with high power density, the adsorbed gas molecules will fly away from the surface of monolayer $WS_2$. Even though adsorption also occurs during this stage, we can still image that the number of desorbing molecules is much larger than that of adsorbing. This is readily understood by considering that the dense pre-adsorbed molecules can fly away directly and rapidly, while the new-adsorbed molecules have to fly from thin air. Consequently, the concentration of adsorbed molecules will decrease while the free electrons will increase, resulting in the conversion from $A^0$ to $A^-$ and the quenching of PL, as $t_1$ to $t_2$ shown in Fig. 2(a). The increased power density will improve the heat effect and thus increase the activation energy. According to the Arrhenius law, the desorption rate will increase synchronously. Thus, PL quenching will become more significantly under higher power density, as PL trajectories presented in Fig. 5(a).

(III) When the power density of irradiation laser is tuned from high to low (such as $P_2$ to $P_1$ shown in Fig. 2(a)), the desorption will be switched off while the adsorption may still survive. In the case, the molecule concentration adsorbed on the $WS_2$ surface will increase, and thus the free electrons will be depleted. Consequently, the conversion from $A^-$ to $A^0$ happens and PL will restore, as $t_4$ to $t_5$ shown in Fig. 2(a). Similarly, the recovery process becomes obvious as the power density increases. However, PL may stand at a plateau level or even quench again, when the power density is too high that the desorption process is activated again, as the triangle and hexagon shown in Fig. 5(b).

By using the pseudo-first-order model to describe the adsorption/desorption process where adsorption/desorption rate is directly proportional to the difference between equilibrium and current molecule concentration over time, $N(t)$.[36] As a consequent, $N(t)$ can be expressed as (see ESI for detailed derivation):

$$N(t) = N(0) \times e^{-k_{de}(P_{Laser}) \cdot t} \quad (2)$$

$$N(t) = N(0) - (N(0) - N'(0)) \times e^{-k_{ad}(P_{Laser}) \cdot t} \quad (3)$$

Equation 2 describes the desorption process, and equation 3 gives the adsorption process (the formula for the combination of the two processes has also provided in the ESI), where $N(0)$ is saturation concentration, $N'(0)$ is the initial concentration of the recovery process (*i.e.*, the molecule concentration at $t_4$ in Fig. 2(a)). $k_{de}(P_{Laser})$ and $k_{ad}(P_{Laser})$ are the rate constant of desorption and adsorption at a certain power density, respectively. According to the Arrhenius law, the higher the power density (thus the more pronounced the laser-induced heat effect), the larger the rate constants. As the solid lines shown in Fig. 5(a) and 5(b), PL quenching and recovery processes can be well described by equations 2 and 3, respectively. The determined rate constants have been presented in Fig. 5(d) and 5(e), respectively. As we expected, both desorption and adsorption rates increase as the power density increases. Furthermore, the desorption rate at the power density of 900 kW/cm$^2$ is $5.4 \times 10^{-2}$ s$^{-1}$, which is in good agreement with $5.8 \times 10^{-2}$ s$^{-1}$, the conversion rate from A$^0$ to A$^-$ (as presented in Fig. 3(c)). The adsorption rate at the power density of $P_1$ also agrees reasonably well with the conversion rate from A$^-$ to A$^0$ ($9.8 \times 10^{-3}$ s$^{-1}$ *vs.* $7.7 \times 10^{-3}$ s$^{-1}$). These results strongly support our assumption.

3.4 *Discussions*

Our adsorption and desorption model can be further verified by putting the sample under a blowing $N_2$ atmosphere and a vacuum chamber. As expected, the reversible manipulation is fully lost in these two conditions. At first glance, our model has explored all the results successfully; however further experiments are still necessary to understand this reversible manipulation. Firstly, we attribute the PL quenching and recovery process to the physical desorption or adsorption of gas molecules (such as $O_2$ and $H_2O$) and utilize the activation energy to explore the results at different power densities. While physical adsorption is generally treated as a non-activated process with a rapid rate (even instantaneous), thus the origin of the activation energy assumed here is still elusive. Secondly, the exact values of activation energy for both adsorption and desorption are still unknown, although the activation energy of desorption higher than that of adsorption has been assumed in the context. These two values might be uncovered by further theoretical calculations and experimental designs, for example, performing this reversible manipulation under different temperature conditions. Thirdly, although equations 2 and 3 can fit the results perfectly up most occasions, some of them still deviate from this model slightly, such as the quenching curve at the power density of 900 kW/cm$^2$ and the recovery curve at the power density of 20 kW/cm$^2$ shown in Fig. 5(a) and 5(b), respectively. The bi-exponential functions are more reasonable than equations 2 and 3 (see Fig. S4 in ESI for the detailed comparison). This deviation hints that more complicated reactions have occurred in this reversible manipulation. Finally, although promising applications have been proposed and reversible micropatterning on monolayer $WS_2$ has been presented in principle (Fig. 2(d)-2(e)), some more practical and conceptual devices are still in design.

## 4. Conclusion

In conclusion, we have shown an all-optical manipulation on PL of monolayer $WS_2$ that the conversion between neural excitons and negative trions can be readily and reversibly fulfilled by changing the power density of irradiation laser. The conversion rates between excitons and trions have been determined by analyzing PL intensity and spectral changes during the manipulation. A potential model, attributing the reversible manipulation to the laser-assisted adsorption and desorption of gas molecules that will deplete or release free electrons from the surface of $WS_2$, has been proposed and explored the experimental results perfectly. The conversion rates between excitons and trions in good agreement with the adsorption/desorption rates further confirms our model. Our finds not only enable nondestructive, reversible, quantitative control of PL emission of monolayer $WS_2$ without electrostatic gating but also provide an all-optical manipulation at desired locations on layered 2D materials with high spatial resolution. These features enable promising applications in micropatterning, optical data storage, and display technology.

**Electronic Supporting Information.**

The Electronic Supporting Information (ESI) is available free of charge on the website, including experimental setup, PL evolution in the initial stage, PL trajectory during the quenching process, the derivation of the adsorption/desorption model, and the comparison of the experimental results with different models.

**Notes:**

The authors declare no competing financial interest.

**Acknowledgment.**

The authors gratefully acknowledge support from the National Key Research and Development Program of China (Grant No. 2017YFA0304203), National Natural Science Foundation of China (Grant No. 11434007), Natural Science Foundation of China (Nos. 61875109, 61527824, 61675119, 11504216, and 61605104), PCSIRT (No. IRT_17R70), 1331KSC and 111 project (Grant No. D18001).

**References:**

1. X. Duan, C. Wang, A. Pan, R. Yu and X. Duan, *Chem. Soc. Rev.*, 2015, **44**, 8859-8876.
2. Q. H. Wang, K. Kalantar-Zadeh, A. Kis, J. N. Coleman and M. S. Strano, *Nat. Nanotechnol.*, 2012, **7**, 699-712.
3. K. F. Mak, C. Lee, J. Hone, J. Shan and T. F. Heinz, *Phys. Rev. Lett.*, 2010, **105**, 136805.
4. Y. Lee, G. Ghimire, S. Roy, Y. Kim, C. Seo, A. K. Sood, J. I. Jang and J. Kim, *ACS Photon.*, 2018, **5**, 2904-2911.
5. D. Sun, Y. Rao, G. A. Reider, G. Chen, Y. You, L. Brezin, A. R. Harutyunyan and T. F. Heinz, *Nano Lett.*, 2014, **14**, 5625-5629.
6. L. Yuan and L. Huang, *Nanoscale*, 2015, **7**, 7402-7408.
7. K. F. Mak, K. He, C. Lee, G. H. Lee, J. Hone, T. F. Heinz and J. Shan, *Nat. Mater.*, 2013, **12**, 207-211.
8. J. S. Ross, S. Wu, H. Yu, N. J. Ghimire, A. M. Jones, G. Aivazian, J. Yan, D. G. Mandrus, D. Xiao, W. Yao and X. Xu, *Nat. Commun.*, 2013, **4**, 1474.
9. S. Mouri, Y. Miyauchi and K. Matsuda, *Nano Lett.*, 2013, **13**, 5944-5948.
10. P. K. Chow, R. B. Jacobs-Gedrim, G. Jian, L. Toh-Ming, Y. Bin, T. Humberto and K. Nikhil, *ACS Nano*, 2015, **9**, 1520-1527.
11. T. Korn, S. Heydrich, M. Hirmer, J. Schmutzler and C. Schüller, *Appl. Phys. Lett.*, 2011, **99**, 102109.
12. X. Fan, W. Zheng, H. Liu, X. Zhuang, P. Fan, Y. Gong, H. Li, X. Wu, Y. Jiang, X. Zhu, Q. Zhang, H. Zhou, W. Hu, X. Wang, X. Duan and A. Pan, *Nanoscale*, 2017, **9**, 7235-7241.
13. M. S. Kim, S. J. Yun, Y. Lee, C. Seo, G. H. Han, K. K. Kim, Y. H. Lee and J. Kim, *ACS Nano*, 2016, **10**, 2399-2405.
14. P. Namphung, Y. Weihuang, S. Jingzhi, S. Xiaonan, W. Yanlong and Y. Ting, *ACS Nano*, 2014, **8**, 11320.
15. M. W. Iqbal, M. Z. Iqbal, M. F. Khan, M. A. Kamran, A. Majid, T. Alharbi and J.

Eom, *RSC Adv.*, 2016, **6**, 24675-24682.

16. S. Feng, C. Cong, N. Peimyoo, Y. Chen, J. Shang, C. Zou, B. Cao, L. Wu, J. Zhang, M. Eginligil, X. Wang, Q. Xiong, A. Ananthanarayanan, P. Chen, B. Zhang and T. Yu, *Nano Research*, 2018, **11**, 1744-1754.

17. A. K. M. Newaz, D. Prasai, J. I. Ziegler, D. Caudel, S. Robinson, R. F. Haglund Jr and K. I. Bolotin, *Solid State Commun.*, 2013, **155**, 49-52.

18. S. Tongay, J. Zhou, C. Ataca, J. Liu, J. S. Kang, T. S. Matthews, L. You, J. Li, J. C. Grossman and J. Wu, *Nano Lett.*, 2013, **13**, 2831-2836.

19. W. He, C. Qin, Z. Qiao, G. Zhang, L. Xiao and S. Jia, *Carbon*, 2016, **109**, 264-268.

20. C. Qin, Z. Qiao, W. He, Y. Gong, G. Zhang, R. Chen, Y. Gao, L. Xiao and S. Jia, *J. Mater. Chem. C*, 2018, **6**, 2329-2335.

21. F. Cadiz, C. Robert, G. Wang, W. Kong, X. Fan, M. Blei, D. Lagarde, M. Gay, M. Manca, T. Taniguchi, K. Watanabe, T. Amand, X. Marie, P. Renucci, S. Tongay and B. Urbaszek, *2D Mater.*, 2016, **3**.

22. H. Van Ngoc, Y. Qian, S. K. Han and D. J. Kang, *Sci. Rep.*, 2016, **6**, 33096.

23. Y. Gong, Z. Lin, G. Ye, G. Shi, S. Feng, Y. Lei, A. L. Elías, N. Perea-Lopez, R. Vajtai and H. Terrones, *ACS Nano*, 2015, **9**, 11658.

24. W. Zhao, Z. Ghorannevis, K. K. Amara, J. R. Pang, M. Toh, X. Zhang, C. Kloc, P. H. Tan and G. Eda, *Nanoscale*, 2013, **5**, 9677-9683.

25. E. Del Corro, A. Botello-Mendez, Y. Gillet, A. L. Elias, H. Terrones, S. Feng, C. Fantini, D. Rhodes, N. Pradhan, L. Balicas, X. Gonze, J. C. Charlier, M. Terrones and M. A. Pimenta, *Nano Lett.*, 2016, **16**, 2363-2368.

26. M. R. Molas, K. Nogajewski, M. Potemski and A. Babinski, *Sci. Rep.*, 2017, **7**, 5036.

27. A. Berkdemir, H. R. Gutiérrez, A. R. Botello-Méndez, N. Perea-López, A. L. Elías, C.-I. Chia, B. Wang, V. H. Crespi, F. López-Urías, J.-C. Charlier, H. Terrones and M. Terrones, *Sci. Rep.*, 2013, **3**.

28. M. Currie, A. T. Hanbicki, G. Kioseoglou and B. T. Jonker, *Appl. Phys. Lett.*, 2015, **106**.

29. H. Y. Nan, Z. L. Wang, W. H. Wang, Z. Liang, Y. Lu, Q. Chen, D. W. He, P. H. Tan,

F. Miao, X. R. Wang, J. L. Wang and Z. H. Ni, *ACS Nano*, 2014, **8**, 5738-5745.

30. A. Venkatakrishnan, H. Chua, P. Tan, Z. Hu, H. Liu, Y. Liu, A. Carvalho, J. Lu and C. H. Sow, *ACS Nano*, 2017, **11**, 713-720.

31. Z. Qiao, C. Qin, W. He, Y. Gong, G. Zhang, R. Chen, Y. Gao, L. Xiao and S. Jia, *Opt. Express*, 2017, **25**, 31025-31035.

32. B. Zhu, X. Chen and X. Cui, *Sci. Rep.*, 2015, **5**, 9218.

33. A. A. Mitioglu, P. Plochocka, J. N. Jadczak, W. Escoffier, G. L. J. A. Rikken, L. Kulyuk and D. K. Maude, *Phys. Rev. B*, 2013, **88**.

34. H. M. Oh, G. H. Han, H. Kim, J. J. Bae, M. S. Jeong and Y. H. Lee, *ACS Nano*, 2016, **10**, 5230-5236.

35. V. Orsi Gordo, M. A. G. Balanta, Y. Galvao Gobato, F. S. Covre, H. V. A. Galeti, F. Iikawa, O. D. D. Couto, F. Qu, M. Henini, D. W. Hewak and C. C. Huang, *Nanoscale*, 2018, **10**, 4807-4815.

36. Y. Gong, C. Qin, W. He, Z. Qiao, G. Zhang, R. Chen, Y. Gao, L. Xiao and S. Jia, *RSC Adv.*, 2017, **7**, 53362-53372.

**Figure captions**

**Fig. 1** Characterizations of the prepared sample. Optical image (a) and atomic force microscopy (AFM) image of $WS_2$ prepared by chemical vapor deposition (CVD). The inset is the height profile of the selected line. (c) Raman spectroscopy of the prepared $WS_2$ excited by 532 nm. The vibrational modes for prominent peaks have been assigned. (d) Photoluminescence (PL) spectra of monolayer $WS_2$, which is fit to Lorentzians (green is the exciton component, $A^0$; orange is the trion component, $A^-$).

**Fig. 2**. (a) PL trajectory of monolayer $WS_2$ varied as laser irradiation with different power densities. In this set of experiments, $P_1$ and $P_2$ are 20 kW/cm² and 900 kW/cm², respectively. (b-e) PL images of monolayer $WS_2$ under the laser excitation of $P_1$ at time $t_0$, $t_4$, $t_5$, and $t_6$, respectively. Scale bar: 5 μm.

**Fig. 3** Time evolution of PL spectra of monolayer $WS_2$ obtained during the quenching process with the excitation of 532 nm and the power density of 900 kW/cm². (a) PL spectra as a function of laser irradiation duration from $t_1$+5 s to $t_1$+90 s, as labeled in Fig. 2a. (b) Representative PL spectra (normalized to the maximum PL intensity) under four different laser irradiation durations. All the spectral profiles are deconvoluted into two peaks (neutral exciton, $A^0$, and negative trion, $A^-$) using Lorentzian curves. (c) Spectral weights and (d) peak energies of $A^0$ and $A^-$ as a function of laser irradiation durations, respectively. The bottle panel presents the binding energy of trion, $\Delta(E_{A0}-E_{A-})$, varied as laser irradiation during the quenching processing.

**Fig. 4** Time evolution of PL spectra obtained during the recovery process with the excitation of 532 nm and the power density of 20 kW/cm$^2$. (a) PL spectra as a function of laser irradiation duration from $t_4$ to $t_4$+340 s. (b) Representative PL spectra (normalized to the maximum PL intensity) under four different laser irradiation durations. All the spectral profiles are deconvoluted into two peaks. (c) Spectral weights and (d) peak energies of $A^0$ and $A^-$ as a function of laser irradiation durations, respectively. The bottle panel presents the binding energy of trion varied as laser irradiation during the recovery processing.

**Fig. 5** PL evolution and the proposed mechanism. Normalized PL trajectories of monolayer WS$_2$ under different power densities in the quenching (a) and recovery (b) processes. Solid lines are the fitting results according to equations 2 and 3, respectively. (c) Schematic of the proposed mechanism. Laser-induced heat effect will assist the adsorption and desorption of gas molecules from the surface of monolayer WS$_2$, resulting in the depletion and release of free electrons. The energy barriers during the adsorption and desorption processes have been schematically illustrated. (d) and (e) are the fitted desorption and adsorption rates derived from the solid lines in (a) and (b), respectively.

**Figure_1**

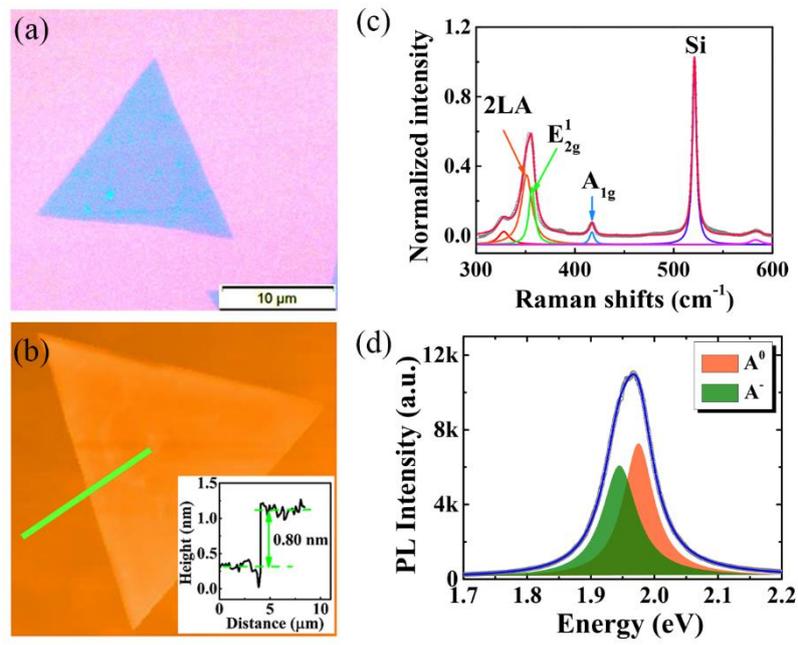

**Figure_2**

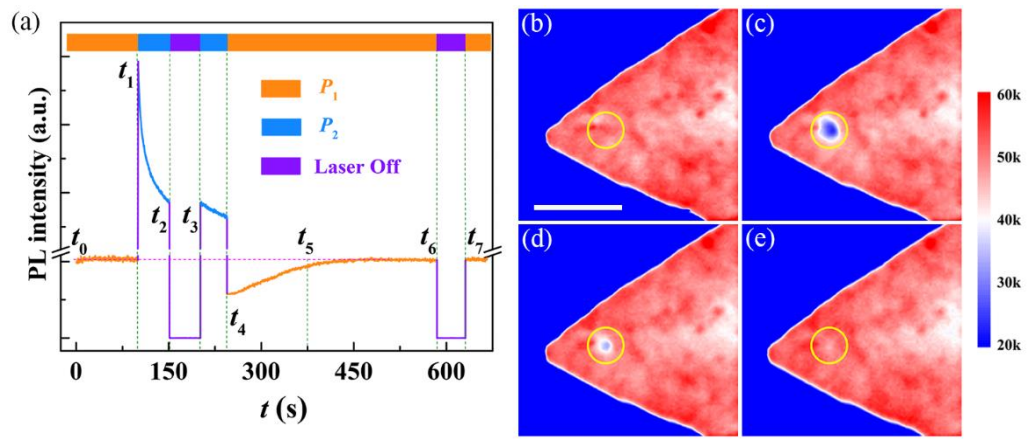

**Figure_3**

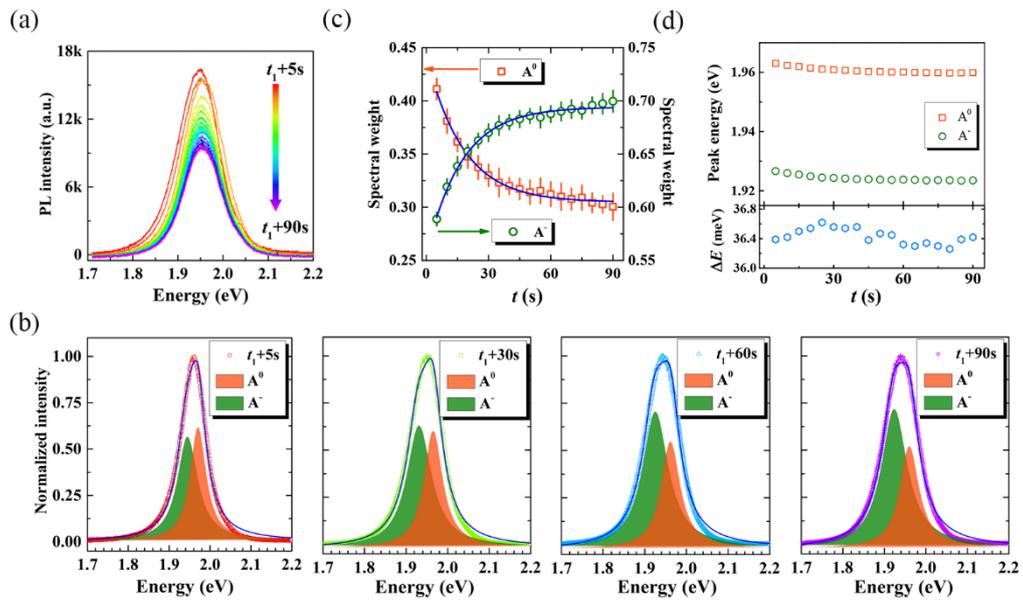

**Figure_4**

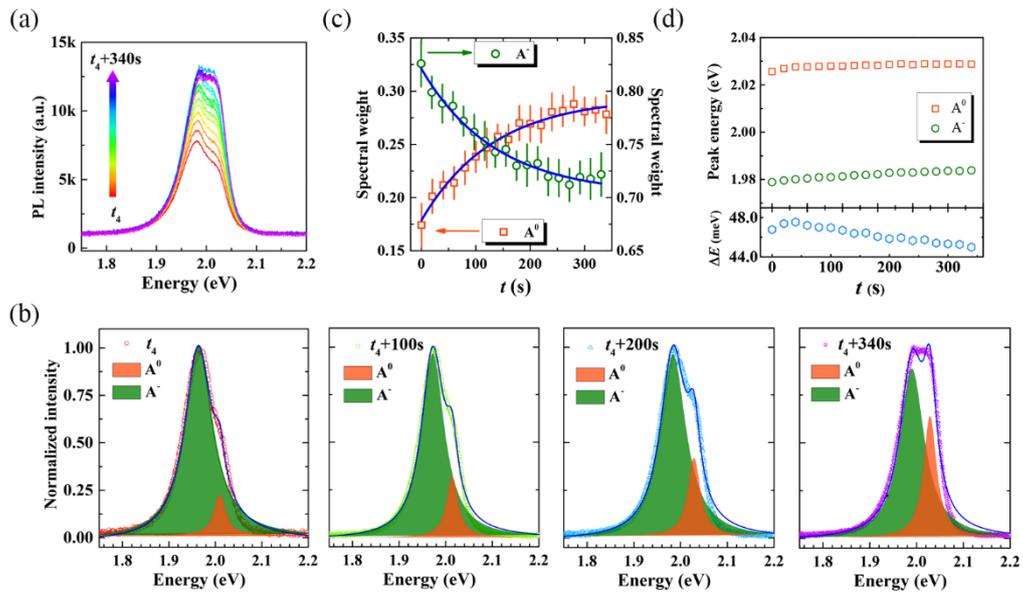

**Figure_5**

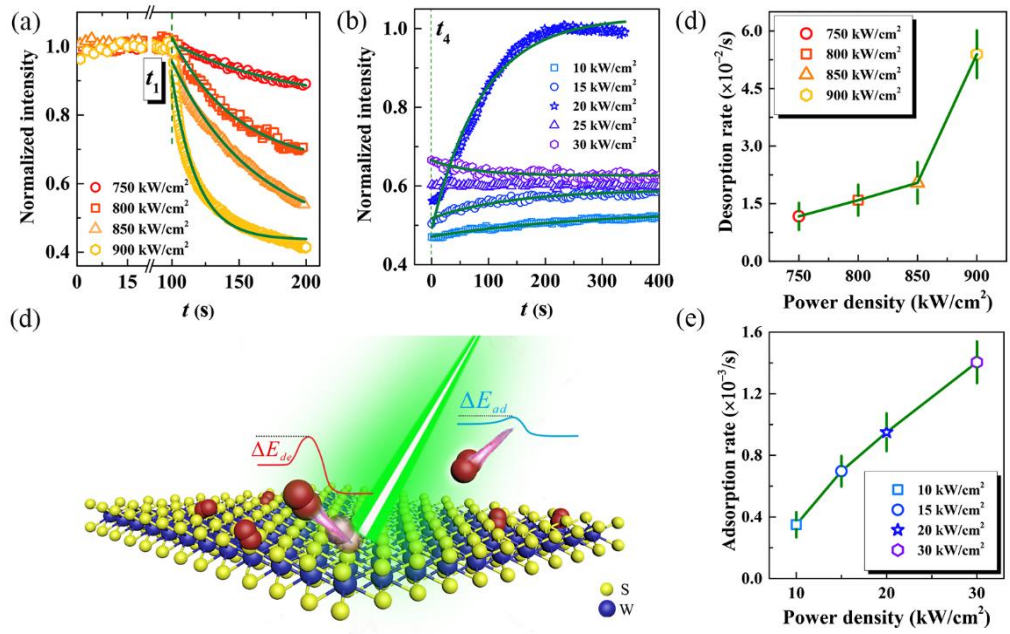